\documentclass[paper,notoc]{JHEP3}

\author{\footnote{KRYSTHAL collaboration} L.A. Harland-Lang$^1$, V.A. Khoze$^{2,3}$, M.G. Ryskin$^{2,4}$, W.J. Stirling$^{1,2}$\\ 
  $^1$Cavendish Laboratory, University of Cambridge,
  J.J.\ Thomson Avenue, Cambridge, CB3 0HE, UK\\
  $^2$ Department of Physics and Institute for Particle Physics Phenomenology, University of Durham, DH1 3LE, UK\\
$^3$ School of Physics \& Astronomy, University of Manchester,
Manchester M13 9PL, UK
$^4$ Petersburg Nuclear Physics Institute, Gatchina, St. Petersburg, 188300, Russia}

\title{Central exclusive quarkonium production with tagged forward protons at RHIC}

\abstract{We study the Central Exclusive Production (CEP) of $\chi_{cJ}$ and $\eta_c$ mesons at RHIC in proton-proton collisions. We consider the $\chi_{cJ}\to J/\psi\,+\,\gamma$ decay channels and, recalling that the $J_z=0$ suppression of the $J=1,2$ states can be compensated by their larger branchings to $J/\psi\,+\,\gamma$, present predictions of rates and distributions
for $\chi_{c(0,1,2)}$ production. Particular attention is paid to the impact of $p_\perp$ cuts applied to the outgoing protons, which can influence the relative rates significantly. The distribution in the azimuthal angle difference between the outgoing protons and the proton $p_\perp^2$ is also studied, and shown to depend sensitively on the spin and parity of the centrally produced meson, as well as being affected by the soft survival factors, $S^2$.  Two- and four-body decays, which are particularly relevant for $\chi_{c0}$ production, are also considered. We show that in the two-body case, backgrounds from `direct' QCD production based on both perturbative and non-perturbative models are expected to be under control.}

\preprint{IPPP/10/87\\ DCPT/10/174 \\ Cavendish-HEP-10/19}

\usepackage{subfigure}
\usepackage{multirow}
\usepackage{helvet}
\usepackage{amsmath}
\usepackage{amssymb}
\usepackage{setspace}
\usepackage{setspace}
\usepackage[dvips]{graphicx}
\usepackage{epsfig}
\def\lesim{ \;\raisebox{-.7ex}{$\stackrel{\textstyle <}{\sim}$}\; }
\def\be{\begin{equation}}
\def\ee{\end{equation}}

\begin{document}

\section{Introduction}

Central exclusive production (CEP) processes in high-energy collisions represent a very promising and novel way to study both QCD and new physics at hadron-hadron colliders. In particular, these reactions provide a valuable tool to investigate in detail the properties of resonance states, from `old' SM mesons to BSM Higgs bosons (see for instance \cite{close}--\cite{Albrow:2010zz}). The CEP of an object $X$ may be written in the form
\begin{equation}\nonumber
pp({\bar p}) \to p \,+\, X\,+\, p({\bar p})\,
\end{equation}
where `+' signs are used to denote the presence of large rapidity gaps. An attractive advantage of these reactions is that they provide an especially clean environment in which to measure the nature and quantum numbers (in particular, the spin and parity) of the centrally produced state $X$, see for example \cite{HKRSTW}, \cite{Kaidalov03}~-~\cite {BSM}. This is particularly true if the outgoing proton momenta can be measured by forward proton taggers: the measured proton distributions give  spin-parity information about the state $X$ and probe the models of soft diffraction used to calculate the rapidity gap survival factors, ${\rm S}^2$~\cite{Kaidalov03,HarlandLang10}.

Recently there has been much activity in the study of central diffractive processes both theoretically (for recent references see, for example, \cite{Khoze04}--\cite{mps}) and experimentally at the Tevatron~\cite{CDFgg}--\cite{d0}, by selecting events with large rapidity gaps separating the centrally produced state from the dissociation products of the incoming protons. In particular, central exclusive $\gamma\gamma$ \cite{CDFgg}, dijet \cite{CDFjj,d0} and $\chi_c$ \cite{Aaltonen09} production have been successfully observed at the Tevatron.
 As noted in~\cite{HarlandLang10,Khoze04gg}, these can serve as standard candle processes with which we can check our predictions for new physics CEP at the LHC. Indeed, the observed rates of all three CEP processes measured at the Tevatron are in broad agreement with theoretical expectations \cite{KMRprosp,Khoze04,Khoze04gg,HarlandLang09,Khoze00a}, which lends credence to the overall theoretical framework and motivates further investigation of new and SM CEP physics at the Tevatron, LHC and, as we shall emphasise here, RHIC. Of special interest is the CEP of heavy quarkonia since it could provide important information on the physics of bound states and can in particular test the current ideas and methods of QCD, such as effective field theories and lattice QCD.

A new area of experimental studies of CEP with tagged forward protons at c.m.s energies up to 500~GeV is now being explored by the STAR Collaboration at RHIC \cite{Guryn08}--\cite{meson}. A capability to trigger on and to measure the outgoing forward protons provides an excellent means to extend the physics reach in studying CEP processes in exceptionally clean conditions. The encouraging preliminary results collected in 2009 during Phase I are already available~\cite{meson} and, hopefully, the large data sample expected from the measurements in Phase II \cite {LeeDIS}--\cite{meson} should provide some very interesting exclusive physics results. Motivated by this, we discuss in this paper the potential for observing exclusive charmonium ($\chi_{cJ}$ and $\eta_c$) production at RHIC with tagged forward protons (we note that the proton momentum loss $\xi$ acceptance covers the mass range $M_X \sim \xi \sqrt{s}$ of these states~\cite{LeeDIS}), paying particular attention to the new and interesting information that the forward proton distributions can provide. We note that the roman pot (RP) detectors at RHIC are uniquely positioned for observing the CEP of such states with tagged protons, as these measurements will not be possible at other hadron-hadron colliders in the foreseeable future.

This paper is organised as follows. In Section~\ref{theory} we review the theory of the CEP of a massive object $X$ in proton-proton collisions.
In Section~\ref{results}  we present predictions for rates and distributions corresponding to $\chi_{cJ}$ and $\eta_c$ production, paying particular attention to 
distributions in the azimuthal angle difference between the outgoing protons and in the proton transverse momenta. Two- and four-body $\chi_c$ 
decays could be experimentally important observables, and so in Section~\ref{pipi} we discuss the corresponding non-resonant backgrounds. 
Finally, Section~\ref{conclusions} contains a summary of our results and some further observations. 

\section{Theory}\label{theory}
\begin{figure}[b]
\begin{center}
\includegraphics[scale=0.9]{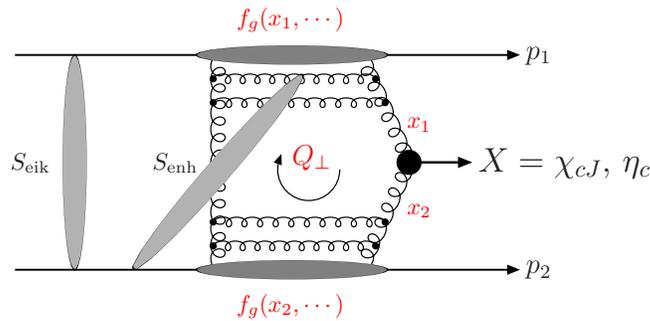}
\caption{The perturbative mechanism for the exclusive process $pp \to p\,+\, X \, +\, p$, with the eikonal and enhanced survival factors 
shown symbolically.}
\label{fig:pCp}
\end{center}
\end{figure} 
The formalism used to calculate the perturbative $\chi_c$ CEP cross section is explained in detail elsewhere~\cite{Kaidalov03,HarlandLang10,Khoze00a,Khoze00} and so we will only review the relevant aspects here. The amplitude is described by the diagram shown in Fig.~\ref{fig:pCp} where the hard subprocess $gg \to X$ is initiated by gluon-gluon fusion and the second $t$-channel gluon is needed to screen the colour flow across the rapidity gap intervals. We can write the `bare' amplitude in the factorised form~\cite{Khoze04,KKMRext}
\begin{equation}\label{bt}
T=\pi^2 \int \frac{d^2 {\bf Q}_\perp\, \mathcal{M}}{{\bf Q}_\perp^2 ({\bf Q}_\perp-{\bf p}_{1_\perp})^2({\bf Q}_\perp+{\bf p}_{2_\perp})^2}\,f_g(x_1,x_1', Q_1^2,\mu^2;t_1)f_g(x_2,x_2',Q_2^2,\mu^2;t_2) \; ,
\end{equation}
where the $f_g$'s in (\ref{bt}) are the skewed unintegrated gluon densities of the proton: in the kinematic region relevant to CEP, they are given in terms of the conventional (integrated) densities $g(x,Q_i^2)$. $t_i$ is the 4-momentum transfer to proton $i$ and $\mu$ is the hard scale of the process, taken typically to be of the order of the central particle mass: we use $\mu=M_X/2$ in what follows. $\mathcal{M}$ is the colour-averaged, normalised sub-amplitude for the $gg \to X$ process
\begin{equation}\label{Vnorm}
\mathcal{M}\equiv \frac{2}{M_X^2}\frac{1}{N_C^2-1}\sum_{a,b}\delta^{ab}q_{1_\perp}^\mu q_{2_\perp}^\nu V_{\mu\nu}^{ab} \; .
\end{equation}
Here $a$ and $b$ are colour indices, $M_X$ is the central object mass, $V_{\mu\nu}^{ab}$ represents the $gg \to X$ vertex and $q_{i_\perp}$ are the transverse momenta of the incoming gluons, given by
\begin{equation}
q_{1_\perp}=Q_\perp-p_{1_\perp}\,, \qquad
q_{2_\perp}=-Q_\perp-p_{2_\perp}\,,
\label{qperpdef}
\end{equation}
where $Q_\perp$ is the momentum transferred round the gluon loop while $p_{i_\perp}$ are the transverse momenta of the outgoing protons. Only one transverse momentum scale is taken into account in (\ref{bt}) by the prescription
\begin{align}\nonumber
Q_1 &= {\rm min} \{Q_\perp,|({\bf Q_\perp}-{\bf p}_{1_\perp})|\}\;,\\ \label{minpres}
Q_2 &= {\rm min} \{Q_\perp,|({\bf Q_\perp}+{\bf p}_{2_\perp})|\} \; .
\end{align}
Explicitly, the $gg\to\chi_c,\eta_c$ vertices are given by~\cite{HarlandLang09}
\begin{align}\label{V0}
&V_0=\sqrt{\frac{1}{6}}\frac{c}{M_\chi}((q_{1_{\perp}}q_{2_{\perp}})(3M_\chi^2-q_{1_{\perp}}^2-q_{2_{\perp}}^2)-2q_{1_{\perp}}^2q_{2_{\perp}}^2)  \; ,
\\ \label{V1}
&V_1=-\frac{2ic}{s} p_{1,\nu}p_{2,\alpha}((q_{2_\perp})_\mu(q_{1_\perp})^2\!-\!(q_{1_\perp})_\mu(q_{2_\perp})^2)\epsilon^{\mu\nu\alpha\beta}\epsilon^{*\chi}_\beta  \; ,\\ \label{V2}
&V_2=\frac{\sqrt{2}cM_\chi}{s}(s(q_{1_\perp})_\mu(q_{2_\perp})_\alpha+2(q_{1_\perp}q_{2_\perp})p_{1\mu}p_{2\alpha})\epsilon_\chi^{*\mu\alpha}  \; ,
\\ \label{V0m}
&V_{0^-}=ic_\eta(q_{1_\perp}\times q_{2_\perp})\cdot n_0\;,
\end{align}%
where the normalisation factors are given by
\begin{align}\nonumber
c_{\chi_c}&=\frac{1}{2\sqrt{N_C}}\frac{16 \pi \alpha_S(\mu^2)}{(q_1 q_2)^2}\sqrt{\frac{6}{4\pi M_{\chi_c}}}\phi'_P(0), \;,\\
c_{\eta_c}&=\frac{1}{\sqrt{N_C}}\frac{4 \pi \alpha_S(\mu^2)}{(q_1 q_2)}\frac{1}{\sqrt{\pi M_{\eta_c}}}\phi_S(0)\;,
\end{align}
and $\phi_{S(P)}(0)$ is the $S(P)$-wave wavefunction at the origin. 
The renormalisation scale $\mu^2$ is assumed to be of the order of $m_c^2$. 
The differential cross section at $X$ ($=\chi_c$, $\eta_c$) rapidity $y_X$ is then 
\begin{equation}\label{ampnew}
\frac{{\rm d}\sigma}{{\rm d} y_X}\!=\!\langle S^2_{\rm enh}\rangle\!\!\int\!\!{\rm d}^2\mathbf{p}_{1_\perp} {\rm d}^2\mathbf{p}_{2_\perp}\! \frac{|T(\mathbf{p}_{1_\perp},\mathbf{p}_{2_\perp}))|^2}{16^2 \pi^5} S_{\rm eik}^2(\mathbf{p}_{1_\perp},\mathbf{p}_{2_\perp})\; ,
\end{equation}
where $T$ is given by (\ref{bt}) and $S^2_{\rm eik}$ is the `eikonal' survival factor, calculated using a generalisation of the `two-channel eikonal' model for the elastic $pp$ amplitude (see ~\cite{KMRtag} and references therein for details). For comparison, we will also consider in Section~\ref{results} a simple one-channel approach, where the elastic amplitude is described by a single Pomeron exchange, and is taken to have the Gaussian form
\begin{equation}\label{app}
A_{pp}(s,k_t^2)=is\,C^{*}\sigma^{\rm tot}_{\rm pp}(s)\exp(-B k_t^2/2)\;,
\end{equation}
where we take $C^*=1.3$ to account for the possibility of proton excitations ($p\to N^*$) in intermediate states~\cite{Kaidalov:1971ta}, and we use $\sigma^{\rm tot}_{\rm pp}=60$ mb and $B=14.7 \,{\rm GeV}^{-2}$ at $\sqrt{s}=500$ GeV~\cite{KMRsoft}.

Besides the effect of eikonal screening $S_{\rm eik}$, there is some suppression caused by the rescatterings of the intermediate partons (inside the unintegrated gluon distribution $f_g$). This effect is described by the so-called enhanced Reggeon diagrams and usually denoted as $S^2_{\rm enh}$, see Fig.~\ref{fig:pCp}. The value of $S^2_{\rm enh}$ depends mainly on the transverse momentum of the corresponding partons, that is on the argument $Q^2_i$ of $f_g(x,x',Q^2_i,\mu^2)$ in (\ref{bt}), and depends only weakly on the $p_\perp$ of the outgoing protons (which formally enters only at NLO). While $S^2_{\rm enh}$ was previously calculated using the formalism of~\cite{Ryskin09}, we now use a newer version of this model~\cite{Ryskintba} which includes the continuous dependence on $Q^2_i$ and not only three `Pomeron components' with different `mean' $Q_i$. We therefore include the $S_{\rm enh}$ factor inside the integral (\ref{bt}), with $\langle S^2_{\rm enh}\rangle$ being its average value integrated over $Q_\perp$.


The expected $\chi_c$ cross section and final-state particle distributions (in particular of the outgoing protons) are therefore determined by a non-trivial convolution of the hard amplitude $T$ and the soft survival factor $S^2_{\rm eik}$. This is modelled in the SuperCHIC Monte Carlo~\cite{SuperCHIC}, which can now explicitly perform the integral (\ref{ampnew}) on an event-by-event basis. This allows for an exact generation of the predicted distributions of the final-state central particles and outgoing protons, as well as a precise evaluation of the expected cross sections after experimental cuts have been imposed, which will depend on both the specific hard process $gg \to X$ as well as the effect of secondary rescatterings.

We end this section with a brief review of the most important uncertainties that are present in our calculation, for more details see~\cite{HarlandLang10,Khoze04,HarlandLang09}. Firstly, the cross section depends on the conventional gluon densities $g(x,Q^2)$ to the fourth power in a region of low $x$ and $Q^2$, where they are poorly determined. Secondly, we have the uncertainty in the values and energy dependence of the non-perturbative survival factors $S^2_{\rm eik}$, $S^2_{\rm enh}$, which we discuss in part in Section~\ref{results}. Thirdly, the infrared stability of the $Q_\perp$ integral (\ref{bt}) depends on the presence of the hard mass scale $M_X$, and it is not completely clear that the $\chi$ mass is large enough to guarantee this. A fourth related uncertainty comes from the possibility of a sizeable `non-perturbative' contribution to the $\chi_{cJ}$ and $\eta_c$ rate, where the exchanged Pomerons couple directly to the $c/\overline{c}$ quarks of the charmonium state.  In total, if we take for each of the four main sources an uncertainty of order $\sim {}^{\times}_{\div}(2-3)$, than adding the errors in quadrature we roughly estimate the uncertainty in the total cross section prediction to be of order $\sim {}^{\times}_{\div} 5$. 

However, we would argue that, as our calculation gives a predicted value for the Tevatron cross section that it is in good agreement with the data, it should also give fairly reliable estimates for the heavy quarkonium CEP cross sections at RHIC (and LHC) energies. Moreover, observables such as the ratios of the predicted perturbative $\chi_{cJ}$ and $\eta_c$ cross sections depend weakly on the PDF set used and so carry smaller overall uncertainties, although in particular the possibility of sizeable spin-parity dependent nonperturbative contributions means it is difficult to quantify this statement. The perturbative contributions will also receive spin and parity dependent higher-order corrections, which are unknown, see~\cite{HarlandLang10}. In particular, $\chi_{c1}$ CEP, which we recall vanishes for on-mass-shell fusing gluons (that is at LO) can only occur at NLO, where we account for the fusion of off-shell gluons. Within the CEP formalism, which makes use of $k_\perp$ factorisation via the explicitly $q_{i_\perp}$ dependent skewed PDFs in (\ref{bt}), we can in part include this initial gluon off-shellness at LO to give a non-vanishing, but heavily suppressed, $\chi_{c1}$ contribution. However, due to this suppression, the $\chi_{c1}$ CEP cross section is highly sensitive to how this off-shellness is included in the $gg\to \chi_{c1}$ matrix element and it may in general also be highly sensitive to NLO effects, which can also generate off-shell incoming gluons. To be safe, as in~\cite{HarlandLang10} we only keep the leading term in the gluon off-shellness $q_{i_\perp}^2/M_{\chi}^2$ to calculate the $\chi_{cJ}$ and $\eta_c$ CEP rates, but keeping all terms can suppress our predicted $\sigma(\chi_{c1})/\sigma(\chi_{c0})$ ratio by a factor of $\sim 3$ (the relative $\chi_{c2}$ to $\chi_{c0}$ rate is roughly unchanged)\footnote{The issue of gluon off-shellness in the case of $\chi_{c0}$ CEP was first addressed in~\cite{Pasechnik07}.}. In the case of the lower mass $\eta_c$ production, this can suppress the $\sigma(\eta_c)/\sigma(\chi_{c0})$ ratio by roughly a factor of $2$.

Furthermore, the predictions for outgoing proton distributions are also less affected by the uncertainties outlined above, although possible non-perturbative contributions will in general affect the predicted distributions, which we calculate only within the perturbative CEP model. However, we note that in the case of $\chi_{c(0,1)}$ and $\eta_c$ production the proton $p_\perp$ and $\phi$ dependence of the perturbative and non-perturbative unscreened amplitudes are approximately equal (up to corrections of order $\langle p_\perp^2\rangle/\langle Q_\perp^2 \rangle$), as these follow from general symmetry principles~\cite{Kaidalov03,HarlandLang09}. For these states this is therefore not expected to be an important source of uncertainty, although in the $\chi_{c2}$ case a large difference between the non-perturbative and perturbative distributions cannot be ruled out. In~\cite{HarlandLang09} it was in particular found that the perturbative and non-perturbative contributions to the total $\chi_{c0}$ cross section are approximately equal ${\rm d}\sigma^{\rm non pert.} \approx {\rm d}\sigma^{\rm pert.}$ (see~\cite{Khoze04} for details of the model used). Previous calculations~\cite{Peng95,Stein93} suggest that the non-perturbative contributions of the three $\chi_c$ states exhibit a similar hierarchy to the perturbative case and so for simplicity we may assume, as in~\cite{HarlandLang10}, the same relative non-perturbative contribution to the $\chi_{c(1,2)}$ cross section as in the $\chi_{c0}$ case, and for consistency we make the same assumption for $\eta_c$ production.

Finally, we note that while the $gg\to\chi_{c2}$ vertex vanishes for $J_z=0$ fusing gluons in the non-relativistic quarkonium approximation, this will in general receive relativistic corrections which will allow a $J_z=0$ component in the production amplitude~\cite{HarlandLang09,Khoze00a}. Although these corrections are expected to be {\it numerically} small (see for instance~\cite{Berg,Li}) we recall that, due to the $J_z=0$ selection rule, CEP of non-relativistic $\chi_{c2}$ (in the $|J_z|=2$ state) is strongly suppressed ($(\sigma(\chi_{c2})/\sigma(\chi_{c0})<1/40)$), and therefore these corrections may have an effect on the $\chi_{c2}$ rate and distributions.

\section{Cross sections and proton distributions}\label{results}

\begin{table}
\begin{center}
\begin{tabular}{|l|c|c|c|c|}
\cline{1-5}
&$\chi_{c0}$&$\chi_{c1}$&$\chi_{c2}$&$\eta_c$\\
\cline{1-5}
$\frac{{\rm d}\sigma}{{\rm d}y_\chi}(pp\to pp(\chi_c))$&27&0.55&0.34&0.24\\
\cline{1-5}
$\frac{{\rm d}\sigma}{{\rm d}y_\chi}(pp\to pp(J/\psi\,+\,\gamma))$&0.31&0.19&0.067\\
\cline{1-4}
\end{tabular}
\caption{Differential cross section (in nb) at $y=0$ for central exclusive $\chi_{cJ}$ and $\eta_c$ production at $\sqrt{s}=500$ GeV. Also shown is the cross section via the $\chi_c \to J/\psi \gamma$ decay channel.}\label{tcross0}
\end{center}
\end{table}

\begin{table}
\begin{center}
\begin{tabular}{l|c|c|c|c|c|c|c|}
\cline{2-6}
&&$\chi_{c0}$&$\chi_{c1}$&$\chi_{c2}$&$\eta_c$\\
\cline{1-6}
\multicolumn{1}{|c|}{\multirow{3}{*}{Fraction (\%)}}&$p_{i_\perp}<0.4$ GeV &47&17&19&13\\
\cline{2-6}
\multicolumn{1}{|c|}{}&$p_{i_\perp}>0.5$ GeV &8.6&9.9&16&20\\
\cline{2-6}
\multicolumn{1}{|c|}{}&$0.5<p_{1_\perp}<0.7$ GeV, $p_{2_\perp}>0.8$ GeV &1.2&1.0&2.2&1.9\\
\cline{1-6}
\end{tabular}
\caption{MC cross section fractions for $\chi_{cJ}$ and $\eta_c$ CEP at $y=0$ and $\sqrt{s}=500$ GeV, for different cuts on the outgoing proton momenta. $S^2_{\rm eik}$ is calculated using parameter set 1 of the two-channel eikonal model of~\cite{KMRsoft}, which accounts for the first N* resonance excitation in low mass ($p\to N^*$) proton dissociation.}\label{tcross}
\end{center}
\end{table}

In Table~\ref{tcross0} we show the predicted differential cross section for $\chi_{cJ}$ and $\eta_c$ CEP at $\sqrt{s}=500$ GeV, calculated following the formalism outlined in Section~\ref{theory} and~\cite{HarlandLang10}.  We note that the RHIC $\chi_{cJ}$ rate is not expected to be significantly lower than the Tevatron prediction of $35$~nb (which is in good agreement with the CDF measurement~\cite{Aaltonen09} once the $\chi_{c(1,2)}$ contributions have been included). This is due to the `eikonal' and `enhanced' survival factors ($S^2_{\rm eik}$,$S^2_{\rm enh}$), which both increase with decreasing $\sqrt{s}$, and therefore compensate the decrease in cross section coming from the smaller gluon density $xg(x,Q^2)$ probed at RHIC energies. Indeed, the predicted $\eta_c$ rate at RHIC is in fact higher than the Tevatron value of $200$ pb, although the cross section difference is well within the theoretical uncertainties.

As was shown in~\cite{HarlandLang10,HarlandLang09}, the $\chi_{c(1,2)}$ rates are expected to be heavily suppressed relative to the $\chi_{c0}$ rate, due to the near-exact $J_z=0$ selection rule that operates for CEP~\cite{Kaidalov03,Khoze00a}, although this suppression may be compensated by the larger $\chi_{c(1,2)}\to J/\psi \gamma$ branching ratios if $\chi_c$ CEP is observed via this decay channel. While the lower mass $\eta_c$ state of course does not decay to $J/\psi \gamma$, it may be observable via 3- or 4-body hadronic decays, which we discuss below. In Table~\ref{tcross0} we therefore also show the predicted differential cross section for $\chi_c$ CEP at $\sqrt{s}=500$ GeV via the $\chi_c \to J/\psi\gamma$ decay channel, which (with photon reconstruction using a converter) would represent a promising way to observe the CEP of these higher spin states. We show in Table~\ref{tcross} the $\chi_{cJ}$ and $\eta_c$ cross section fractions after realistic experimental cuts have been imposed on the outgoing protons\footnote{Note that the machine constraints may also in general affect the $2\pi$ coverage for the outgoing protons: in particular for RHIC Phase II these could give a further factor of $\sim 1/(0.75)^2$ decrease in the observed $\chi_c$ rate~\cite{meson}.}, to take into account the $p_\perp$ acceptance of the roman pot (RP) detectors. In particular, rows 1 ($p_{i_\perp}<0.4$ GeV) and 2 ($p_{i_\perp}>0.5$ GeV) correspond to the RPs for Phases I (currently in place) and II (to be installed for $\sim$2013) of the STAR pp2pp physics programme, respectively~\cite{meson}. As expected, we can see that by selecting events with small $p_\perp$, the ratio of the expected $\chi_{c0}$ to $\chi_{c(1,2)}$ and $\eta_c$ yields increases by a factor of 2--3: as the outgoing proton $p_\perp$ decreases, the $J_z=0$ selection rule becomes more exact and the higher spin states are more heavily suppressed. For higher proton $p_\perp$ the effect of absorptive corrections becomes more important: although the `bare' unscreened $\chi_{c(1,2)}$ amplitudes favour higher $p_\perp$ values, this effect is less significant once screening effects are included. In particular, the $\chi_{c0}$ state is not significantly suppressed when we select events with high $p_\perp$, although some enhancement of the $\chi_{c2}$ rate is found. On the other hand, in the case of the $\eta_c$, for which we recall that the unscreened amplitude also favours higher $p_\perp$ values, there is over a factor of $2$ enhancement relative to the $\chi_{c0}$.

\begin{figure}
\begin{center}
\includegraphics[trim=30 0 0 0,scale=0.6]{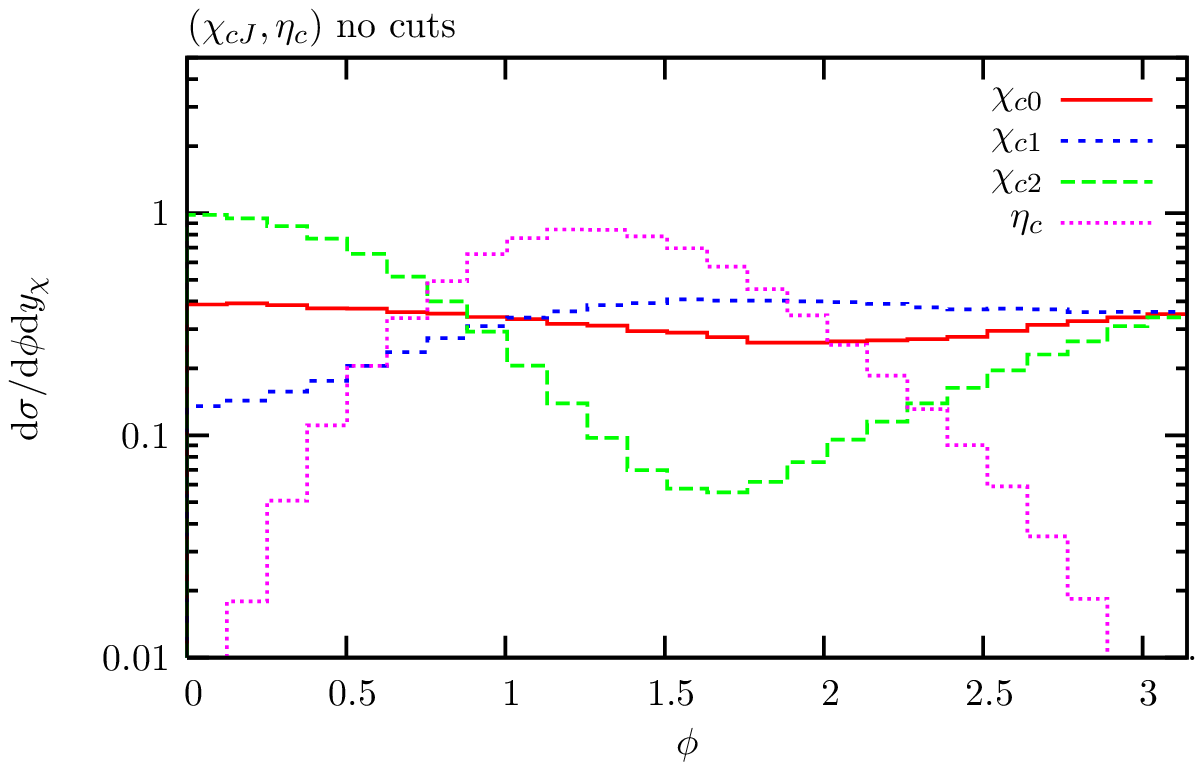}
\includegraphics[scale=0.6]{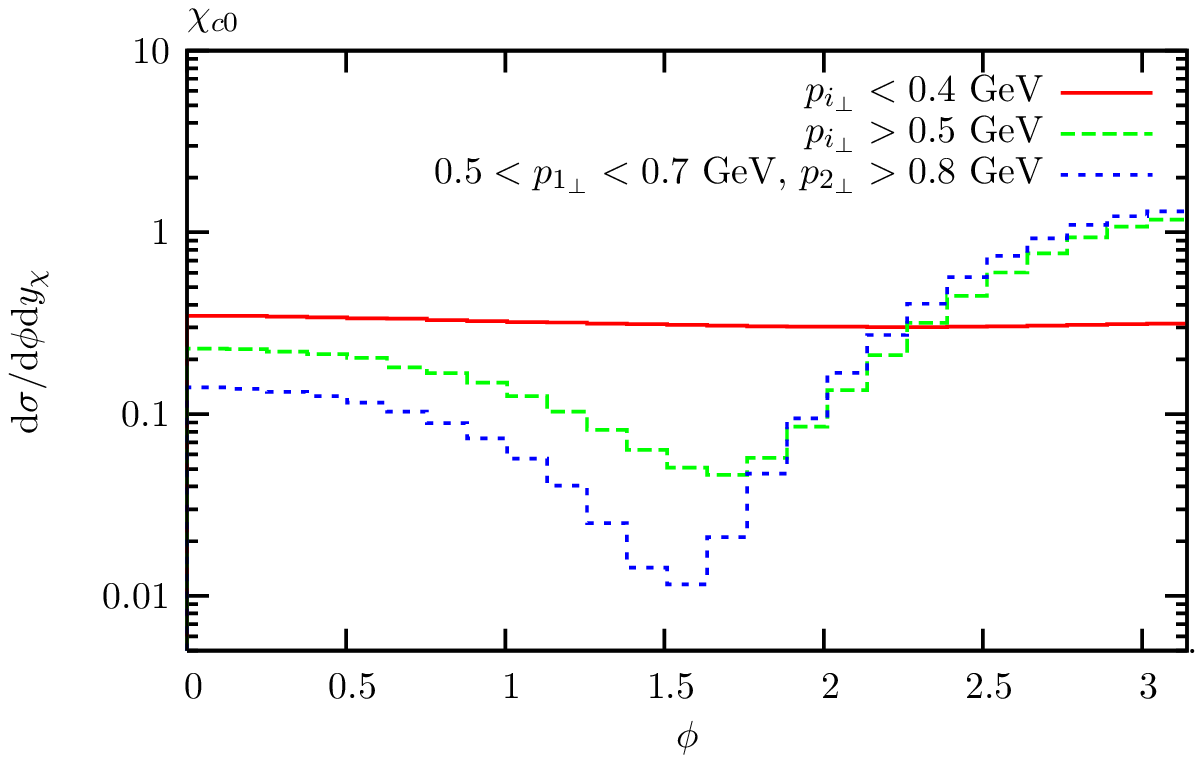}
\includegraphics[trim=30 0 0 0,scale=0.6]{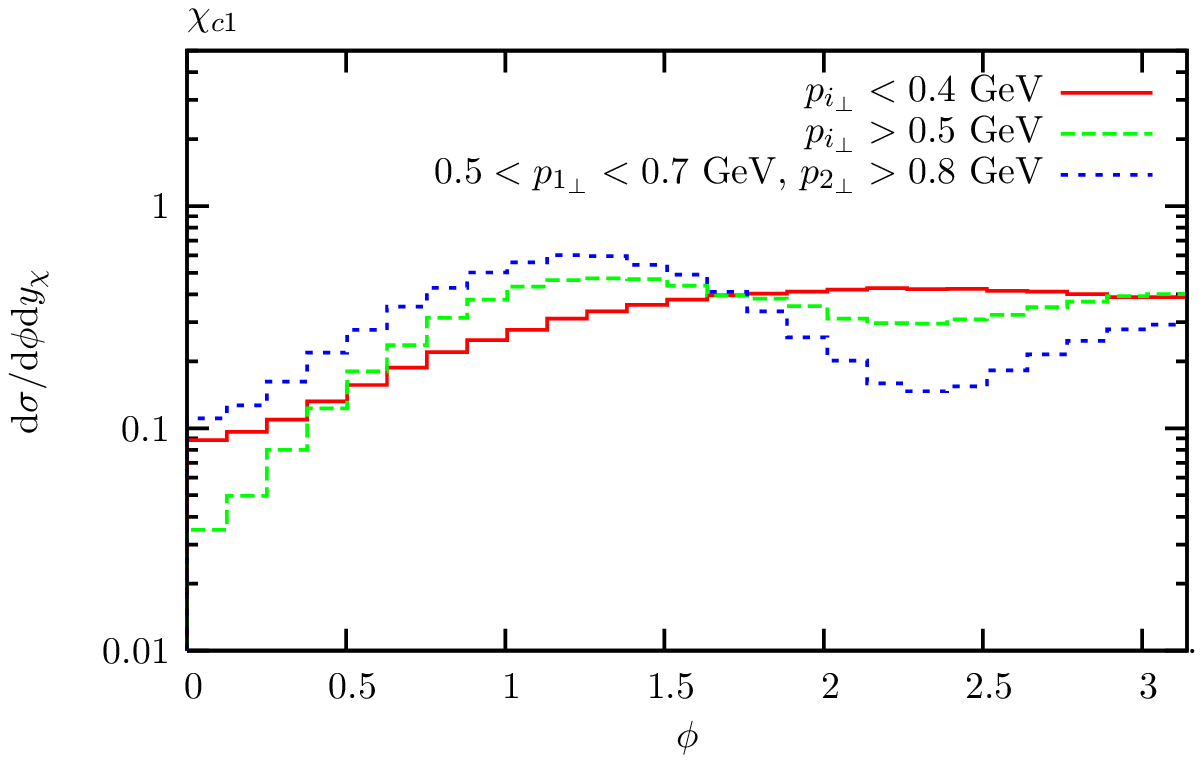}
\includegraphics[scale=0.6]{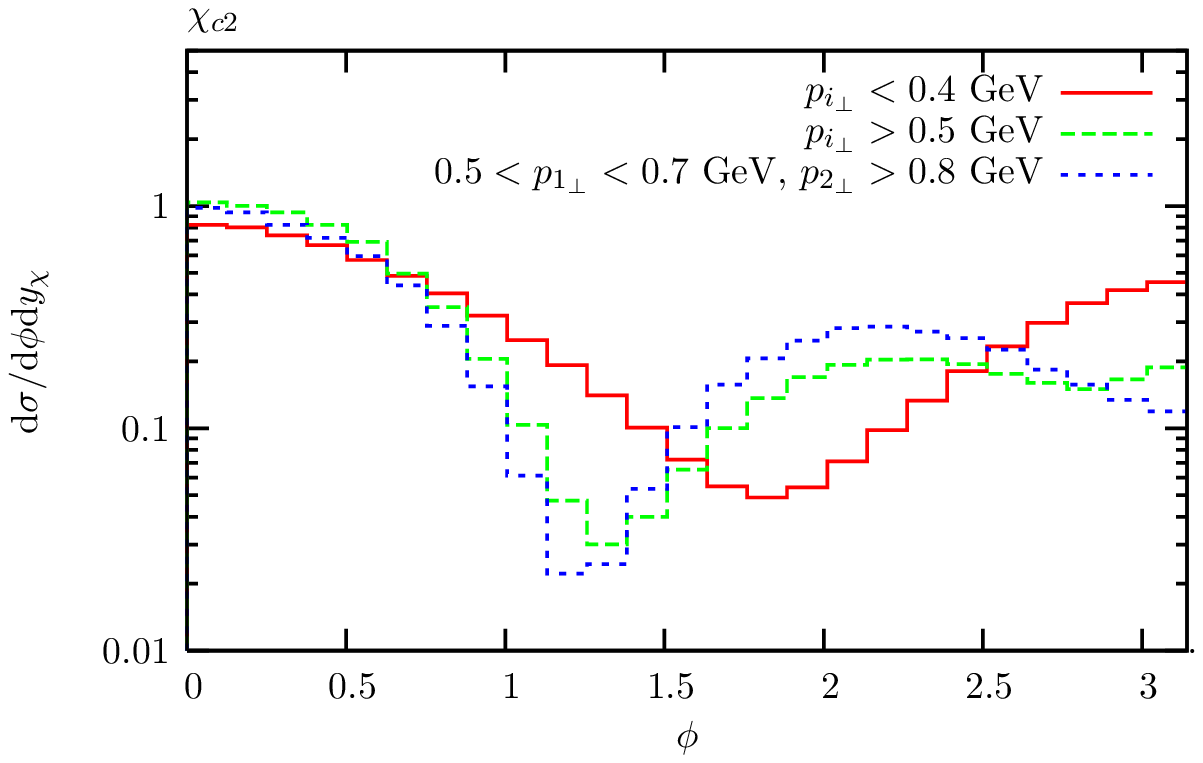}
\includegraphics[scale=0.6]{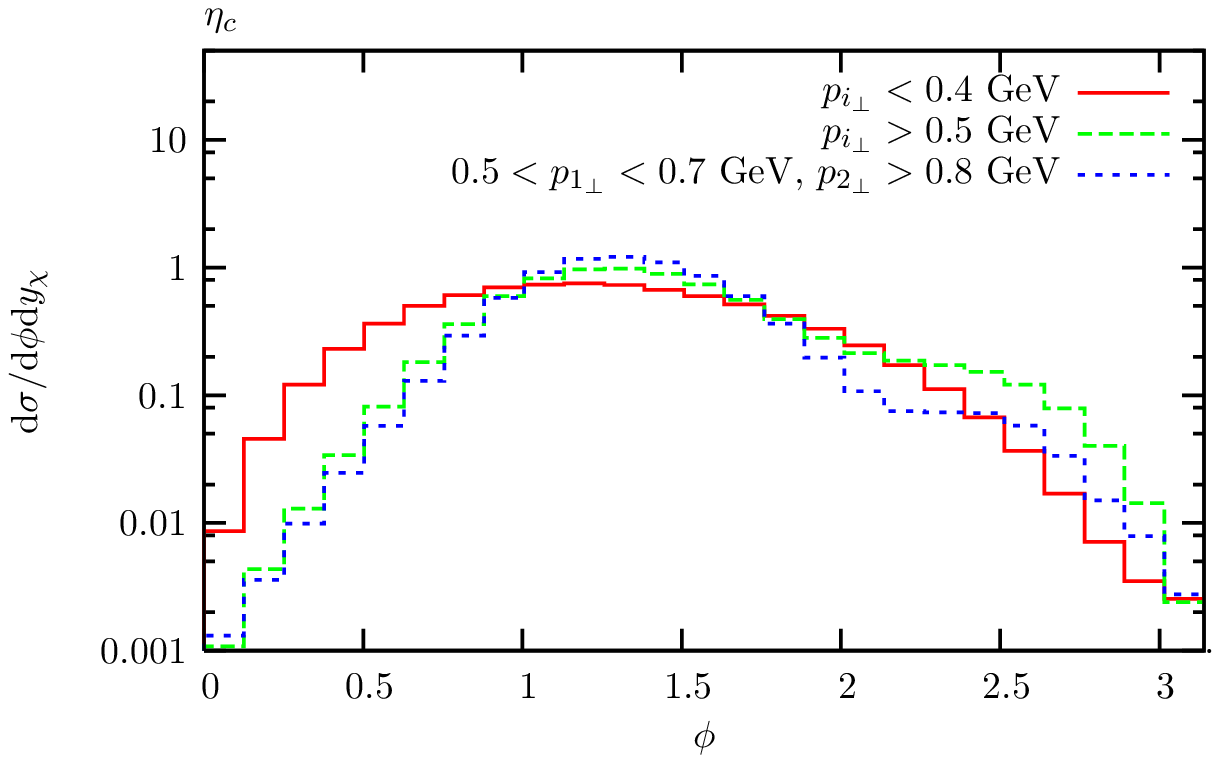}
\caption{Normalised distributions (in arbitrary units) of the difference in azimuthal angle between the outgoing protons for the CEP of $\chi_{c(0,1,2)}$ and $\eta_c$ states at $y=0$ and $\sqrt{s}=500$ GeV, and for a range of cuts on the proton $p_\perp$. $S^2_{\rm eik}$ is calculated using set 1 of the two-channel eikonal model of~\cite{KMRsoft}, which accounts for the first N* resonance excitation in low mass ($p\to N^*$) proton dissociation.}\label{dist1}
\end{center}
\end{figure}

\begin{table}
\begin{center}
\begin{tabular}{|l|c|c|c|c|c|}
\hline
&$\chi_{c0}$&$\chi_{c1}$&$\chi_{c2}$&$\eta_c$\\
\hline
$\langle S^2_{\rm eik}\rangle$, Fit 1&0.092&0.23&0.15&0.26\\
\hline
$\langle S^2_{\rm eik}\rangle$, Fit 2&0.062&0.18&0.11&0.21\\
\hline
$\langle S^2_{\rm eik}\rangle$, Single channel&0.070&0.20&0.13&0.24\\
\hline
\end{tabular}
\caption{`Eikonal' survival factor, $\langle S^2_{\rm eik}\rangle$, averaged over the outgoing proton ${\rm p}_\perp$, calculated using the two-channel eikonal model of~\cite{KMRsoft}, with two different choices of model parameters. Parameter set 1 accounts for the first N* resonance excitation in low mass ($p\to N^*$) proton dissociation, while set 2 includes excitations up to a larger $M^2\sim 6\;{\rm GeV}^2$. Also shown is the result of using the simplified single channel eikonal approach, see (\ref{app}).}\label{tsurv}
\end{center}
\end{table}

\begin{figure}
\begin{center}
\includegraphics[trim=30 0 0 0,scale=0.6]{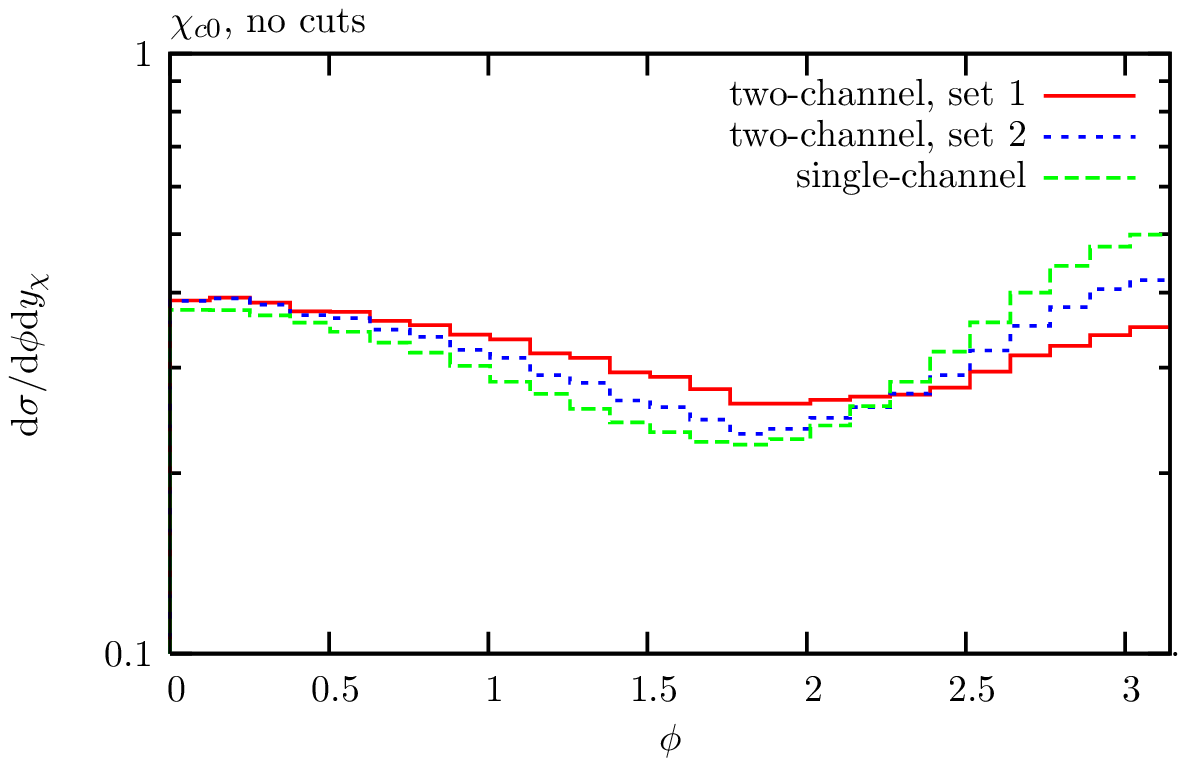}
\includegraphics[scale=0.6]{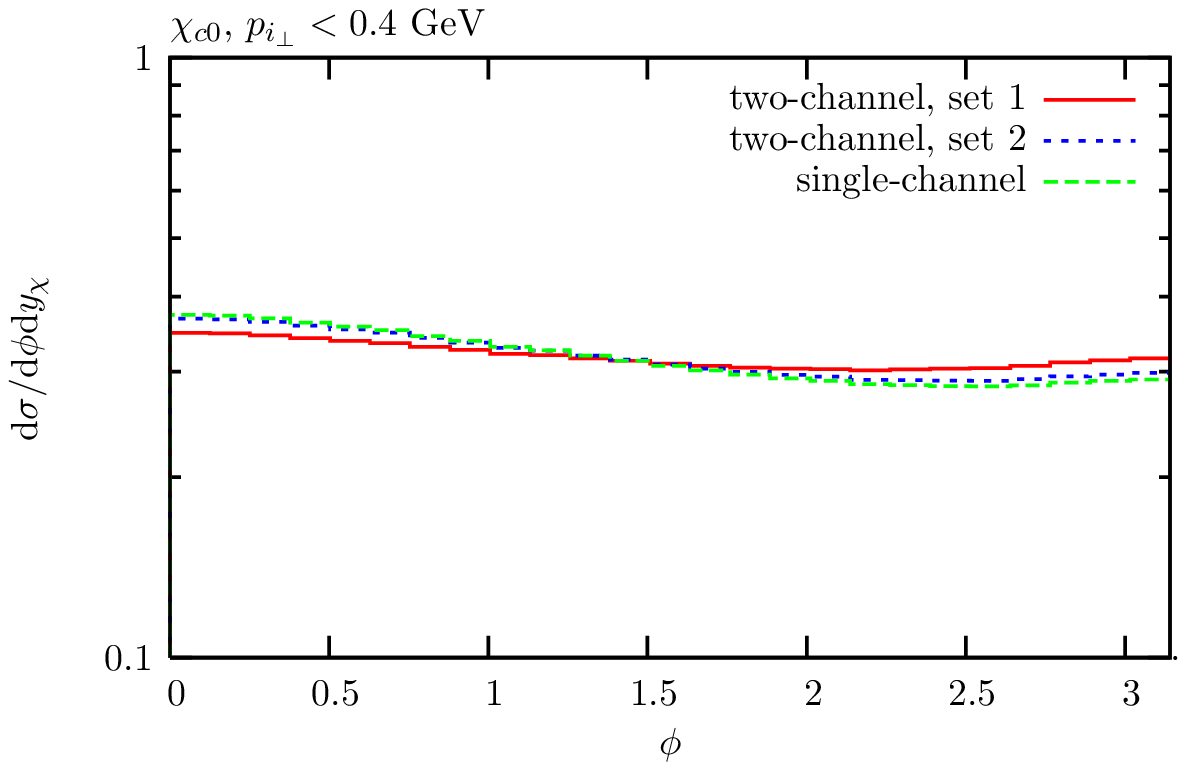}
\includegraphics[scale=0.6]{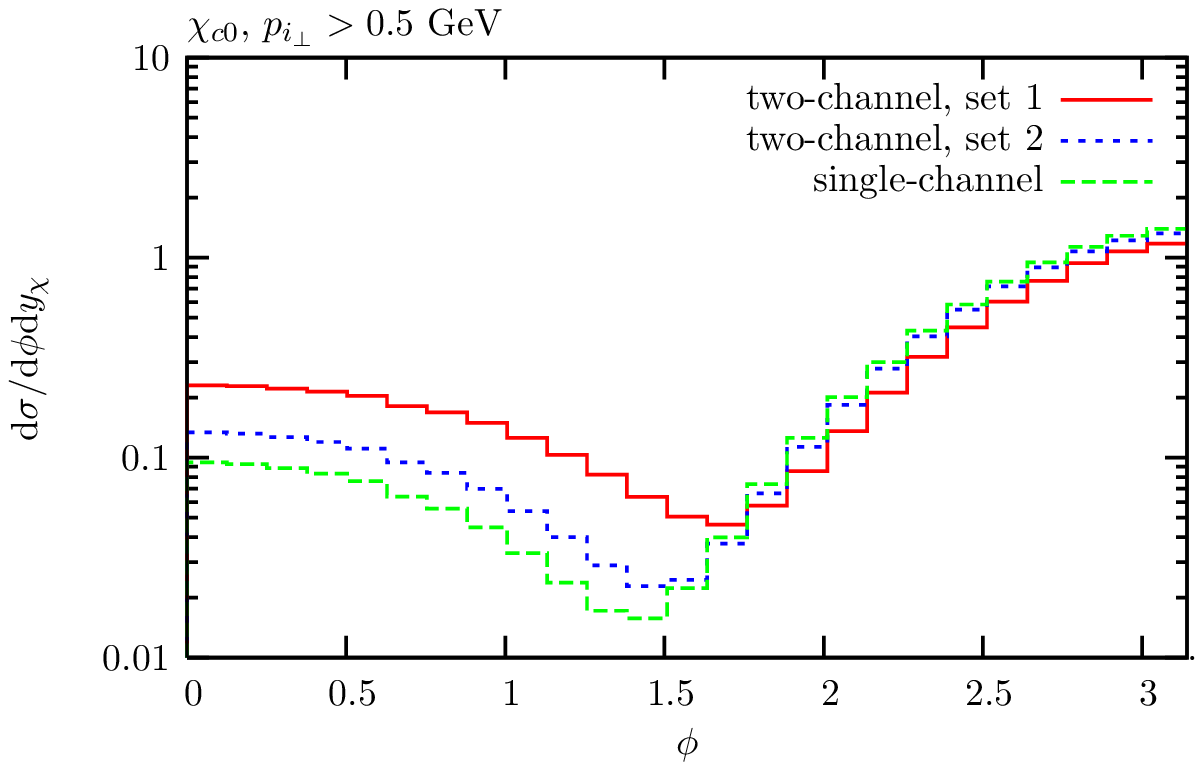}
\caption{Normalised distributions (in arbitrary units) of the difference in azimuthal angle between the outgoing protons for $\chi_{c0}$ CEP at $y=0$, with the survival factor $S^2_{\rm eik}(b_t)$ calculated using the two-channel eikonal model of~\cite{KMRsoft}, with two different choices of model parameters. Parameter set 1 accounts for the first N* resonance excitation in low mass ($p\to N^*$) proton dissociation, while set 2 includes excitations up to a larger $M^2\sim 6\;{\rm GeV}^2$. Also shown is the result of using the simplified single channel eikonal approach, see (\ref{app}). Note that the `two-channel, set 1' $\chi_{c0}$ distributions are the same as those 
plotted in Fig.~\ref{dist1}.} \label{dist2}
\end{center}
\end{figure}

\begin{figure}
\begin{center}
\includegraphics[trim=30 0 0 0,scale=0.6]{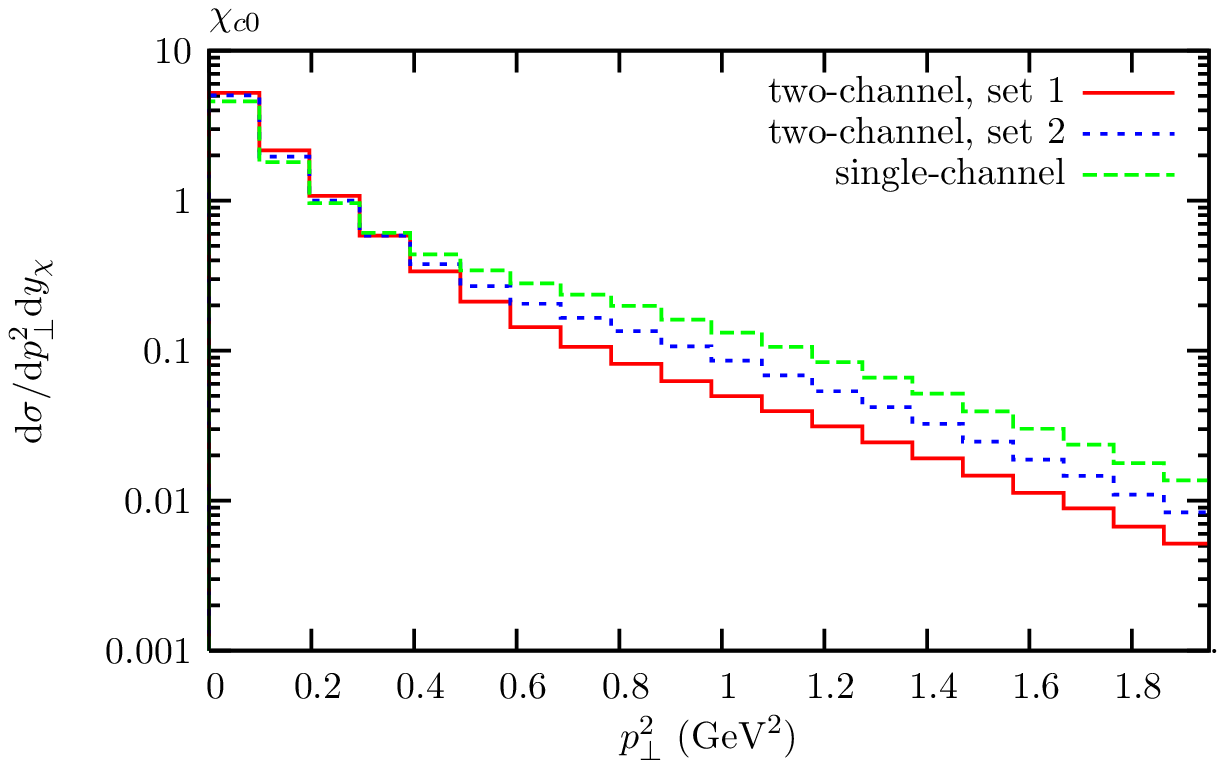}
\includegraphics[scale=0.6]{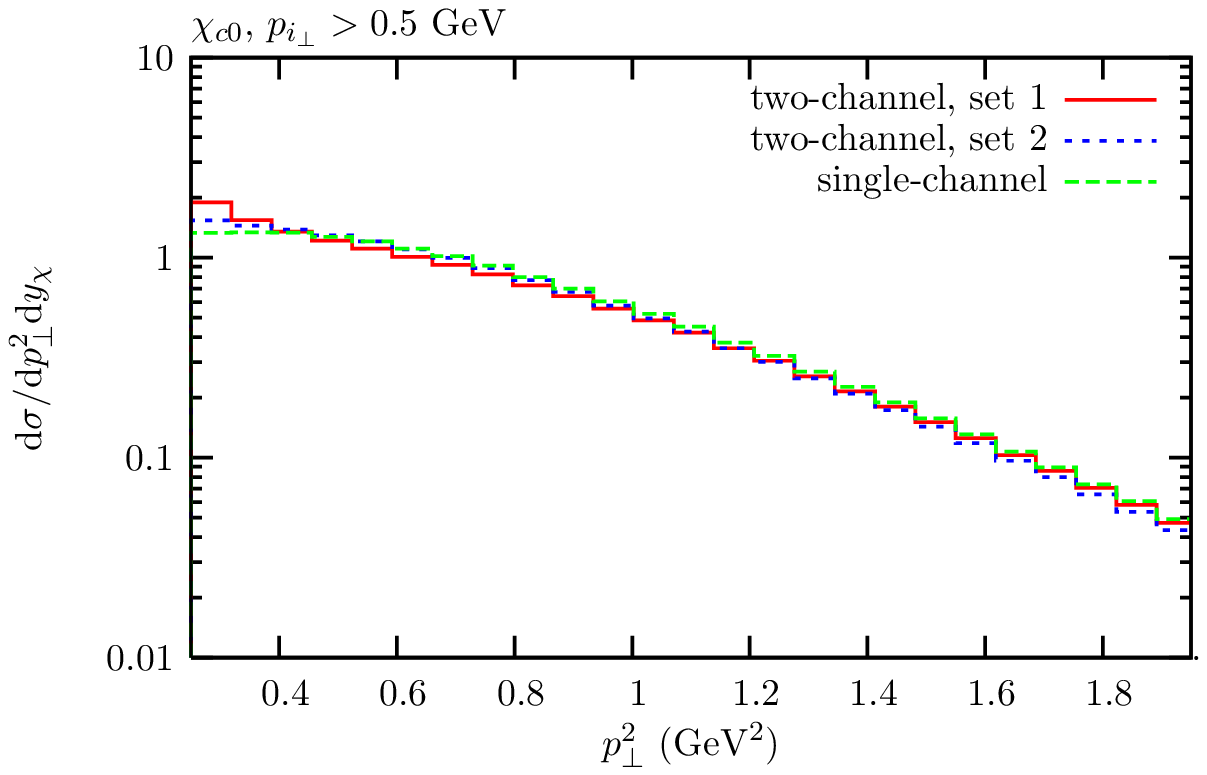}
\includegraphics[scale=0.6]{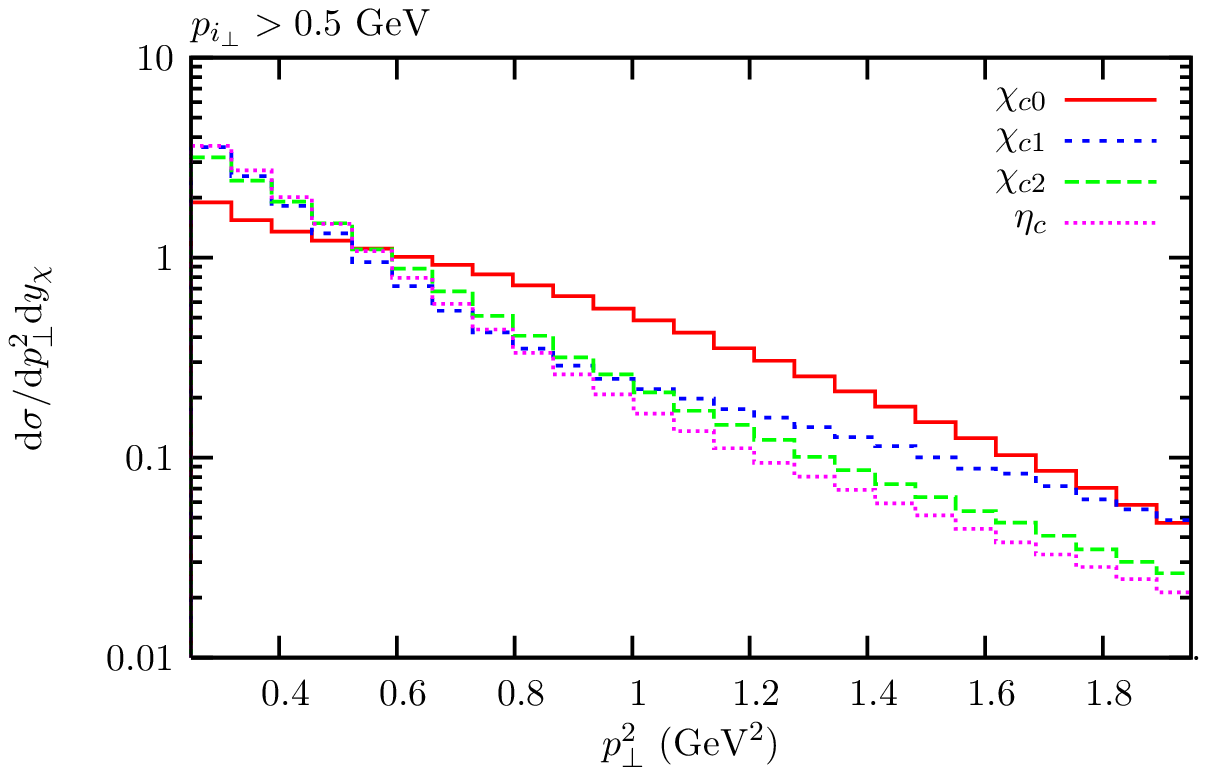}
\caption{Normalised distributions (in arbitrary units) of the $p_\perp^2$ of the outgoing proton, integrated over the other proton $p_\perp$, for $\chi_{cJ}$ and $\eta_c$ CEP at $y=0$, and for $\chi_{c0}$ CEP with the survival factor $S^2_{\rm eik}(b_t)$ calculated using the two-channel eikonal model of~\cite{KMRsoft}, with two different choices of model parameters. Parameter set 1 accounts for the first N* resonance excitation in low mass ($p\to N^*$) proton dissociation, while set 2 includes excitations up to a larger $M^2\sim 6\;{\rm GeV}^2$. Also shown is the result of using the simplified single channel eikonal approach, see Eq.~(\ref{app}). The lower $\chi_{cJ}$ plot uses fit 1.} \label{dist3}
\end{center}
\end{figure}

In Refs.~\cite{Kaidalov03,HarlandLang10}, it was shown that the distributions in $p_\perp$ and $\phi$ of the outgoing protons depend sensitively on the spin and parity of the centrally produced object. This is illustrated in Fig.~\ref{dist1}, where we plot the expected ${\rm d}\sigma/{\rm d}\phi$ distributions for $\chi_{cJ}$ and $\eta_c$ production, calculated using the SuperCHIC MC\footnote{Note that the unscreened $\chi_{c1}$ and $\chi_{c2}$ distributions, shown in~\cite{HarlandLang10}, are different from those presented in the published version of~\cite{Pasechnik:2009qc}. However we have been informed by the authors that their more recent calculation confirms our results~\cite{Roman}.}. By applying different cuts to the outgoing proton $p_\perp$, we can also in principle probe the underlying theory in a more detailed way, and to illustrate this we plot the $\phi$ distributions for three different sets of $p_\perp$ cuts: $p_{(1,2)_\perp}<0.4$ GeV, $p_{(1,2)_\perp}>0.6$ GeV and $0.5<p_{1_\perp}<0.7$ GeV, $p_{2_\perp}>0.8$ GeV. 
For low $p_\perp$ the screening corrections do not change this `bare' behaviour too much, however in the case of a relatively large $p_\perp$ (green and blue lines in Fig.~\ref{dist1}) 
the role of absorptive effects becomes quite visible:
starting from $\phi=0$ the absorptive correction increases with $\phi$ producing a dip in the region of $\phi\sim \pi/2$ for the cases of the $\chi_{c0}$ and $\chi_{c2}$ and about $\phi\sim 2.3$ for the $\chi_{c1}$. We note that these characteristic `diffractive dip' structures have the same physical origin as the proton azimuthal distribution patterns first discussed in~\cite{KMRtag}. For the $\eta_c$ these effects are less significant, although some non-negligible dip structure around $\phi=2-2.5$ can be seen.

As described in~\cite{HarlandLang09}, a further way to extract spin information about the centrally produced $\chi$ state is by measuring the angular distributions of its decays products, in particular here the final state $\mu^+\mu^-$ pair from $J/\psi$ decay. These spin-dependent angular distributions, which are generated by the SuperCHIC MC, would also represent an interesting observable, providing complementary information to the tagged proton distributions.

As well as depending on the spin-parity of the centrally produced particle, the proton distributions will also be affected non-trivially by soft-survival effects~\cite{KMRtag}, in particular the ${\bf p}_\perp$ dependent survival factor $S^2_{\rm eik}({\bf p}_{1\perp},{\bf p}_{2\perp})$, through Eq.~(\ref{ampnew}). We show in Fig.~\ref{dist2} the predicted $\phi$ distribution for $\chi_{c0}$ CEP using the `two-channel' eikonal formalism of~\cite{KMRsoft} for a range of cuts on the proton $p_\perp$. To give some indication of the uncertainty involved in calculating the survival factors, we show the distributions using two different choices of model parameters as well as using the simplified single channel model described in Section~\ref{theory}. In particular, to calculate $S_{\rm eik}^2$ the elastic and $\sigma_{\rm tot}$ $pp$ data were fitted and the available data on low mass ($p\to N^*$) proton dissociation were used, as described in~\cite{KMRsoft,Luna08}. Parameter set 1 and the one-channel enhanced (by $C=1.3$) eikonal model account for the fixed target (mainly old FNAL) data and correspond to the first N* resonance excitation. Set 2 accounts for CERN-ISR data and excitations up to a larger $M^2\sim 6\;{\rm GeV}^2$ (which is correlated with $S^2_{\rm enh}$ based on the KMR model~\cite{Ryskin:2009tk}).

In Table~\ref{tsurv} we show the ${\bf p}_\perp$ averaged suppression factors $\langle S^2_{\rm eik}\rangle$, for the different parameter sets and the simplified single-channel model. Clearly there is some (of order $\lesim 50\%$) non-negligible model dependence in the overall soft suppression factor, which varies depending on the spin-parity of the centrally produced particle. This variation, which results from our incomplete knowledge of the soft amplitude used to calculate the survival factors, represents an important source of uncertainty in any CEP cross-section predictions. On the other hand, the ratios of the $S^2_{\rm eik}$ values between the different $\chi_{cJ}$ and $\eta_c$ states are less model dependent, varying by $\lesssim 15\%$. Moreover, the distributions are not too dependent on the choice of parameter set, although some difference, in particular for larger proton $p_\perp$ values (where we recall the amplitude is more sensitive to soft survival effects), is apparent. This is equally true for the case of the single-channel eikonal model, in particular when experimental cuts are imposed. We note that the $\chi_{c1}$ and $\chi_{c2}$ $\phi$ distributions are also similarly dependent on the choice of parameter set for the survival factors.

Finally, we show in Fig.~\ref{dist3} the proton $p_\perp^2$ distribution (integrated over the other proton $p_\perp$) for $\chi_{c0}$ CEP using the same two sets of parameters and the simplified single-channel model to calculate the survival factors. When no cuts are imposed on the proton $p_\perp$ we can see that there is in principle an observable difference between the $p_\perp^2$ distributions resulting from the different parameter sets, but this is no longer the case when the most relevant cut $p_{i_\perp}>0.5$ GeV is imposed. As with the $\phi$ distributions, we can see that the $p_\perp^2$ distribution (shown for the case $p_{i_\perp}>0.5$ GeV) also depends on the spin-parity of the centrally produced state.

$\chi_{c0}$ CEP, observed via two-body (e.g. $\chi_{c0} \to \pi^+\pi^-$, $K^+K^-$, $p\overline{p}$) or four-body (e.g. $\chi_{c0} \to 2(\pi^+\pi^-)$, $\pi^+\pi^-K^+K^-$) decays also represents an interesting observable at RHIC, provided the direct QCD backgrounds are sufficiently under control --- we will discuss this below. Experimentally, the exceptionally  good resonance mass reconstruction (of order of a few MeV) provided by the excellent charge particle identification and high resolution tracking in the STAR TPC will greatly aid in increasing the S/B ratio. On the other hand, the higher spin $\chi_{c(1,2)}$ states are expected to give negligible contributions via these decay channels. To give an indication of the expected rates, in Table~\ref{brs} we summarise the branching ratios, taken from~\cite{PDG}, for some of the two- and four-body $\chi_{c0}$ decays that may be experimentally relevant at RHIC. Note that around 6\% of all $\chi_{c0}$ decays are to $h^+h^-$ or $2(h^+h^-)$ with $h=\pi,K$. The corresponding mesonic two-body decays ($\pi\pi,KK,\phi\phi,\omega\omega,\eta\eta$ etc.) of the $\chi_{c2}$ are all at the $10^{-3}$ level, while the 4-body decay modes (e.g. $\pi^+\pi^-K^+K^-$, $2(\pi^+\pi^-)$) are at the $1\%$ level~\cite{PDG}, and may also be useful.\footnote{We note that currently the STAR mass acceptance for $\pi^+\pi^-$ and $K^+K^-$ states does not extend beyond $\approx 3.2$~GeV due to particle ID, i.e. below the $\chi_c$ threshold~\cite{meson}. However, even without  particle ID, observation of $\chi_{c0}\to\pi^+\pi^-,K^+K^-$ should be possible: in particular, for each event we can include the possibility in the analysis that the two charged tracks are either $\pi^+\pi^-$ or $K^+K^-$. This will increase the non-resonant background by about a factor of 2, but given the excellent mass resolution in the STAR TPC, a clear $\chi_c$ peak should still be observable. Nevertheless, any future extension of the STAR mass reach (with  particle ID) would certainly improve the experimental situation. Note also that the mass coverage of the $p \bar p$ decay mode is expected to be broader (up to around 4~GeV) due to better particle ID.} In the case of $\eta_c$ production, the three-body (e.g. $K\overline{K}\pi$, with branching $7\%$) and four-body (e.g. direct $2(\pi^+\pi^-)$ decay, with branching $1.2\%$, or via the wide resonance states $\rho^0\rho^0\to 2(\pi^+\pi^-)$, with branching $2\%$) decay modes are the most realistic.

\begin{table}
\begin{center}
\begin{tabular}{|l|c|}
\hline
Mode&Branching ratio\\
\hline
$\pi^+\pi^-$&$(0.56\pm0.03)\times 10^{-2}$\\
\hline
$K^+K^-$&$(0.610\pm0.035)\times 10^{-2}$\\
\hline
$K_S K_S$&$(0.316\pm0.018)\times 10^{-2}$\\
\hline
$\phi\phi$&$(9.2\pm1.9)\times 10^{-4}$\\ 
\hline
$f_0(980)f_0(980)$&$(6.8\pm2.2)\times 10^{-4}$\\ 
\hline
$p\overline{p}$&$(2.28\pm0.13)\times 10^{-4}$\\
\hline
$\Lambda\overline{\Lambda}$&$(3.3\pm0.4)\times 10^{-4}$\\
\hline
$2(\pi^+\pi^-)$&$(2.27\pm0.19)\times 10^{-2}$\\
\hline
$\pi^+\pi^-K^+K^-$&$(1.80\pm0.15)\times 10^{-2}$\\
\hline
\end{tabular}
\caption{Branching ratios for $\chi_{c0}$ two- and four-body decays, taken from Ref.~\cite{PDG}.}\label{brs}
\end{center}
\end{table}

\section{Non-resonant QCD background}\label{pipi}

\begin{figure}
\begin{center}
\subfigure[]{\includegraphics[scale=0.65]{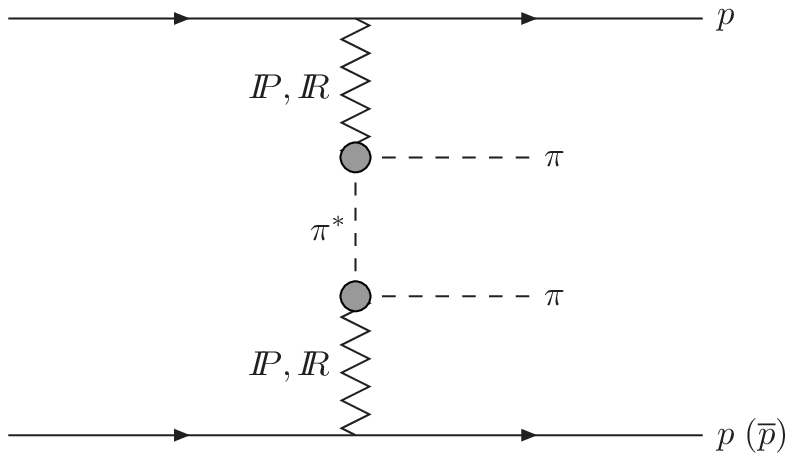}}
\subfigure[]{\includegraphics[scale=1.1]{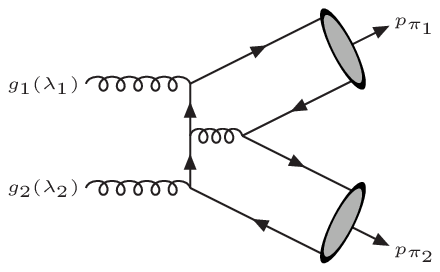}}
\caption{{\bf (a)} Non-perturbative $\pi\pi$ CEP production mechanism -- similar diagrams exist for $K^+K^-$ and $p\overline{p}$ production. {\bf (b)} A typical diagram corresponding to the perturbative $gg\to\pi\pi$ hard subprocess.}\label{pifig}
\end{center}
\end{figure}

In Section~\ref{results} we noted the possibility of observing $\chi_{c0}$ CEP via two-body decay channels, with the $\chi_c \to \pi\pi$ decay being a promising example. We recall that these decay channels, especially $\pi\pi$, $K^+K^-$ and $p\bar{p}$, are ideally suited for spin-parity analysis of the $\chi_c$ states, in particular the fact that the $\chi_{c(1,2)}$ two body branching ratios are in general of the same size or smaller (or even absent for the $\chi_{c1}$) than the $\chi_{c0}$ ensures that the $J_z=0$ selection rule is active, see \cite{Khoze04,HarlandLang09} for more details. However, in this case we may in principle expect a sizeable background resulting from direct production, $pp \to p \,+\,h \bar h \, +\, p$ with $h=\pi,K,p$, and so care must be taken to estimate the expected non-resonant contribution. This can be modelled using a `non-perturbative' framework, mediated by Pomeron-Pomeron fusion with an intermediate off-shell $h=\pi,K,p$ exchanged between the final-state particle pair, see Fig.~\ref{pifig}(a)\footnote{$\pi\pi$ CEP mediated by double Pomeron exchange has been a subject of theoretical studies in the last 40 years or so (see~\cite{aklr,Pumplin:1976dm} for early references and~\cite{ls} for a more recent one). There have also been a variety of experimental results, in particular from the CERN ISR (for recent reviews see \cite{afc,Albrow:2010zz}).}, or via the perturbative 2-gluon exchange mechanism with the $gg\to \pi^+\pi^-,K^+K^-$ coupling modelled using the formalism of~\cite{Brodsky81,Chernyak06}\footnote{We note that $gg\to p\overline{p}$ can also 
in principle be modeled in this way, see~\cite{Lepage79,Lepage80}.}, see Fig.~\ref{pifig}(b). In the kinematic ($M_{\pi\pi}\sim M_\chi$, $p_\perp(\pi)\sim M_\chi/2$) region relevant to $\chi_c$ production we may expect both mechanisms to contribute to the overall cross section. Besides the `bare' amplitudes shown in Fig.~\ref{pifig} we also have to account for the absorptive corrections, that is the eikonal,
$S_{\rm eik}$, and `enhanced', $S_{\rm enh}$, survival factors introduced in Section~\ref{theory}.

Considering first the perturbative mechanism for $\pi^+\pi^-$ production, using purely dimensional arguments we can see that the amplitude to form an exclusive pion pair with large transverse momentum $k_\perp$ will be proportional to the ratio $(f_\pi/k_\perp)^2$ (the pion form factor $f_\pi$ enters through the normalisation of the pion wavefunction $\phi(x)$), that is the cross section of the $gg^{PP}\to \pi^+\pi^-$ hard subprocess contains the numerically small factor $(f_\pi/k_\perp)^4$. Moreover, as we shall discuss in detail in a future publication~\cite{Harlandlangfut}, the LO amplitude for $\pi\pi$ exclusive production in a $J_z=0$ state vanishes in the same way as the $\gamma\gamma\to \pi^0\pi^0$ amplitude~\cite{Brodsky81,Chernyak06}. Thus the process will occur mainly through the admixture of the 
$|J_z|=2$ state of the incoming active gluons $gg^{PP}$, which we recall is strongly suppressed due to the $J_z=0$ selection rule which operates for forward outgoing protons~\cite{Khoze00a}. We note that this result also extends to $K^+K^-$ production by replacing $f_\pi$ by $f_K$ and using a slightly different (asymmetric) wavefunction to account for the non-zero strange quark mass~\cite{Chernyak06}. The net effect is that
the $K^+K^-$  cross section is predicted to be approximately equal to that for $\pi^+\pi^-$ (this has already been observed by BELLE for the case of $\gamma\gamma \to \pi^+\pi^-,K^+K^-$~\cite{Nakazawa04}, see also~\cite{Benayoun89}).

In the case of the non-perturbative contribution, large pion $p_\perp$ values are suppressed by the form factor, $F_\pi(t)$, of the intermediate off-shell pion, although in this case it is not completely clear what form to take for $F_\pi(t)$; as a result there is a large uncertainty in the predicted cross section for this non-perturbative mechanism. In particular, we may either take the `soft' exponential $\exp(bt)$, or the `hard' power-like form $\sim 1/t$, both of which are used in the literature~\cite{Pumplin:1976dm,ls,Desai:1978rh}. As $|t|$ (that is, the pion or kaon $p_\perp$) is increased, these two choices of form factor give vastly different cross sections: while the exponential form gives a negligible cross section, the power-like form gives a comparable cross section to $\chi_c$ CEP. When available, CDF data~\cite{Albrow:2010zz,malbrow} on central exclusive $\pi^+\pi^-$ production in the $\chi_c$ mass range, collected in the same rapidity range ($|\eta| \simeq 1$), would therefore help to further constrain the non-resonance background and may provide useful information about the off-shell pion form factor. However, we note that even taking the power-like pion form factor, which gives the much larger non-resonant cross section estimate, we expect to obtain a background contribution that will be under good control experimentally. In particular, it corresponds to a cross section of about 10-20 pb under the $\chi_c$ resonance mass peak --- an order of magnitude less than the expected $\chi_{c0} \to \pi^+\pi^-$ CEP cross section. This is equally true for the $K^+K^-$ channel. 

For the $p\overline{p}$ decay channel, we again have to consider the production mechanism shown in Fig.~\ref{pifig} (a), but with a proton, rather than pion, line. The Pomeron-proton coupling is about a factor $3/2$ larger than in the pion case, but the proton form factor is steeper (the `hard' power-like form is given by $F_p \sim 1/t^2$). Therefore, for a reasonably large proton $k_\perp>1$ GeV the non-perturbative contribution is expected to be smaller than in the $\pi\pi$-channel.

In the case of 4-body decays, the situation is more complicated. An important part of these signal decays come from the $\chi_c$ decay to two wide resonances, such as $\chi_c \to \rho^0\rho^0\to 2(\pi^+\pi^-)$. An analogous mechanism is possible for the background process, and, taking into account the number of appropriate resonances and the final state interactions, it is hard to give a definite prediction. However, we do not expect the 4-body background to be much larger than for the 2-body channel.

\section{Conclusions}\label{conclusions}

In this paper we have studied Central Exclusive Production of $\chi_{cJ}$ and $\eta_c$ mesons at RHIC, for proton-proton collisions. Because the gap survival factors increase with decreasing $\sqrt{s}$, the production rates at RHIC are in fact expected to be comparable to those at the Tevatron. For the $\chi_{cJ} \to J/\psi + \gamma$ decay channel, the strong suppression of the $J=1,2$ states is compensated by the larger branching ratios for these states. Overall, for this decay channel we predict a cross section $d\sigma/dy_{\chi} \simeq 0.6$~nb at $y_{\chi}=0$, with the $J=0,1,2$ states roughly in the proportion $5:3:1$. We have also considered $\eta_c$ CEP, for which we predict a cross section of approximately $0.2$~nb, which is heavily suppressed relative to the $\chi_{c0}$ (for which the predicted rate is roughly $30$ nb), as we expect from the $J_z^P=0^+$ selection rule. These rates are, however, dependent on the experimental $p_\perp$ cuts on the outgoing protons.  In particular, selecting events with small $p_\perp$ increases the ratio of the $\chi_{c0}$ to $\chi_{c(1,2)}/\eta_c$ yields by roughly a factor of 2: as the outgoing proton $p_\perp$ decreases, the $J_z=0$ selection rule becomes more exact and the higher spin states are more heavily suppressed. SuperCHIC MC~\cite{SuperCHIC}, which generates all of the processes described in this paper with exact proton kinematics, allows for a complete evaluation of the acceptances within the full experimental setup for the different $c\overline{c}$ states.

The distribution in the azimuthal angle difference between the outgoing protons also depends sensitively on the spin and parity of the centrally produced object, and also to some extent on the transverse momentum dependence of the survival factors. While we focused mainly on the $\chi_{cJ} \to J/\psi + \gamma$ decay channel, we also considered two- and four-body decays which are particularly relevant for $\chi_{c0}$ production. We showed that in the two-body case, backgrounds from `direct' QCD production based on both perturbative and non-perturbative models are expected to be under control.

It is worth mentioning that CEP can help to shed light on the nature of the numerous recently discovered charmonium-like mesons ($X, Y, Z$)~\cite{PDG},
and, in particular, on the $X(3872)$ resonance, whose nature is presently unclear~\cite{pahl}. Found in 2003 by the BELLE collaboration~\cite{belle} and then confirmed by CDF~\cite{cdf} and D0~\cite{do} at the Tevatron and by BABAR~\cite{babar}, this narrow state still remains a mystery and, consequently, a subject of intensive theoretical activity (for reviews see for example~\cite {swanson,olsen} and for recent papers~\cite{sim}--\cite{coito} and references therein). As is the case for many of the new charmonium-like states, $X(3872$) cannot be easily fitted into the conventional charmonium spectroscopy, and so a number of more exotic explanations for its origin have been suggested, such as a $DD^*$ molecule, tetraquark, hybrid, some sort of threshold effect, etc. --- see~\cite{swanson,olsen} for more details. However, quite recently there has been a renewal in attempts to interpret it as a regular $1^{++} (2^3P_1)$ charmonium state~\cite{jia}~--~\cite{coito}.\footnote {$2^{-+}$ is another spin-parity assignment allowed by the existing data \cite{CDFjpc}. However as a $1^1D_2$ state within the charmonium interpretation, this looks quite problematic~\cite{jia}~--~\cite{kalash}.}

Additional experimental information on the properties of the $X(3872)$, especially its $J^P$ assignment, could provide new insight into the theoretical understanding of this state. A detailed study of the  CEP process $pp \to p \,+\, X(3872) \, +\, p$ would be a valuable new tool to test the quantum numbers of this state (see \cite{Khoze04}), and so could play an important role in the resolution of the $X(3872)$ puzzle. In particular, a comparison of the proton momentum correlation pattern with the expectations for a $1^{++}$ charmonium state would
provide support for or rule out the interpretation of this resonance as a $2^3P_1$ member of the charmonium family.

Finally, we note that the planned vertex detector will allow STAR to perform a search for the $\chi_{c2}(2P)$ resonance, which has recently been observed at B-factories~\cite{belle1} in the $D\bar{D}$ mode. We do not expect that the CEP rate for this state will be very different, within the theoretical uncertainties, from the prediction given here for the $\chi_{c2}(1P)$ state.

\section{Acknowledgements} 

We thank Mike Albrow, Alan Martin, Risto Orava, Roman Pasechnik, Rainer Schicker,  Antoni Szczurek, and Oleg Teryaev, for useful discussions. We are especially grateful to
Wlodek Guryn for encouraging discussions and information about the pp2pp Roman Pots at the STAR experiment at RHIC.
MGR, LHL and WJS thank the IPPP at the University of Durham for hospitality.
The work by MGR was supported  by the Federal Program of the Russian State RSGSS-3628.2008.2.
This work is also supported in part by the European Community's
Marie-Curie Research Training Network under contract MRTN-CT-2006-035505
`Tools and Precision Calculations for Physics Discoveries at Colliders'
(HEPTOOLS).
LHL acknowledges financial support from the University of Cambridge Domestic Research Studentship scheme.

  B.~Aubert  [The BABAR Collaboration],
  arXiv:1002.0281 [hep-ex].

  \end{document}